\documentclass[11pt]{article}
\usepackage{amssymb}
\usepackage{amsmath}
\usepackage[dvips]{graphicx}
\makeatletter \@addtoreset{equation}{section} \makeatother

\makeatletter
\renewcommand\section{\@startsection {section}{1}{\z@}%
                                   {-5.5ex \@plus -1ex \@minus -.2ex}
                                   {2.3ex \@plus.2ex}%
                                   {\normalfont\large\bfseries}}
\renewcommand\subsection{\@startsection{subsection}{2}{\z@}%
                                     {-3.25ex\@plus -1ex \@minus -.2ex}%
                                     {1.5ex \@plus .2ex}%
                                     {\normalfont\normalsize\bfseries}}
\renewcommand\thesection {\@arabic\c@section}
\renewcommand\thesubsection   {\thesection.\@arabic\c@subsection}
\renewcommand{\@seccntformat}[1]{%
\csname the#1\endcsname.\hspace{1.0em}}
\makeatother

\newcommand{\Tr}{\textrm{Tr}}

\newcommand{\ev}[1]{\langle #1 \rangle}
\newcommand{\myvec}[1]{{\bf #1}} 
\newcommand{\dd}{\textrm{d}}
\newcommand{\LE}{\mathcal{L}_{\rm{E}}}
\newcommand{\SE}{S_{\rm{E}}}
\newcommand{\MSb}{\overline{\rm{MS}}}
\newcommand{\Nc}{N_{\rm c}}
\newcommand{\BG}{B_{\rm{G}}}
\newcommand{\mG}{m_{\rm{G}}}

\begin{document}

\begin{titlepage}
\begin{flushright}
HIP-2006-41/TH\\
\end{flushright}
\begin{centering}
\vfill{}
 
{\Large{\bf Plaquette expectation value and lattice free energy \newline of three-dimensional SU$(\Nc)$ gauge theory}}

\vspace{0.8cm}

A. Hietanen$^{\rm a,b}$, 
A. Kurkela$^{\rm a}$, 

\vspace{0.8cm}

{\em $^{\rm a}$%
Theoretical Physics Division, 
Department of Physical Sciences, \\ 
P.O.Box 64, FI-00014 University of Helsinki, Finland\\ }

{\em $^{\rm b}$%
Helsinki Institute of Physics,\\
 P.O.Box 64, FI-00014 University of Helsinki, Finland\\ }

\vspace*{0.3cm}
 
\noindent
\abstract{
We use  high precision lattice simulations to calculate the plaquette expectation value in three-dimensional SU($\Nc$) gauge theory for $\Nc = 2,3,4,5,8$. 
Using these results, we study the $\Nc$-dependence of the first non-perturbative coefficient in the weak-coupling expansion of hot QCD.
We demonstrate that, in the limit of large $\Nc$, the functional form of the plaquette expectation value with ultraviolet divergences subtracted is $15.9(2)-44(2)/\Nc^2$.
}
\vfill
\noindent
 


\vspace*{1cm}
 
\noindent

\vfill
\end{centering}
\end{titlepage}

%
\section{Introduction}

The determination of QCD pressure up to order $g^6$ is a long-standing
problem in finite-temperature field theory \cite{linde,gpy,gsixg}.
This is the first order where a coefficient of the weak-coupling 
expansion, due to infrared divergences,
gets contributions from an infinite number of loop-diagrams and thus is
non-perturbative. 

However, at high enough temperatures ($T \gtrsim 2 T_c$) the
properties of finite-temperature QCD can be described by dimensionally reduced
effective field theory methods \cite{dr,generic}.
By integrating out temporal degrees of freedom a three-dimensional
pure gauge theory,  called magnetostatic QCD (MQCD), is constructed.
This allows us to isolate all the divergences to MQCD and study
it using lattice calculations. The integration out is most conveniently 
performed perturbatively in $\MSb$ scheme \cite{bn}.

We can relate any lattice regularized quantities within MQCD to the 
continuum scheme
($\MSb$), because MQCD is super-renormalizable. There are ultraviolet  
divergences up to 4-loop level only \cite{farakos}. Terms required 
in the conversion have been determined up to 3-loop level \cite{dir,pt}. 
Infrared divergences cause an additional complication in the
4-loop level. The computation requires an introduction of an IR cutoff,
which then cancels once lattice and $\MSb$ results are
subtracted. This computation has been carried out recently for $\Nc=3$ in
\cite{renzo} using stochastic perturbation theory.

In \cite{plaquette} the plaquette expectation value, which determines 
the non-perturbative contribution, was measured for $\Nc=3$. The
purpose of this paper is to extend the results to study the $\Nc$-dependence
of this observable. We carry out lattice measurements of the plaquette
with $\Nc=2,3,4,5$ and $8$ to obtain the $\Nc$-dependence. We also get an
independent approximation for the $\Nc=3$ result. This acts as a
consistency check for the whole  pressure calculation. Namely, we
expect to see smooth $\Nc$-dependence in the observable.

Additionally, there are various other physical motivations to study the $\Nc$-dependence and
especially the large-$\Nc$ limit of SU($\Nc$) gauge theories \cite{thooft}.
 The limit $\Nc \rightarrow \infty$ simplifies
the theory significantly, but nevertheless the phenomenology is in many
ways similar to SU(3). 
These reasons have motivated numerous large-$\Nc$ limit studies on the
lattice \cite{largen,KorthalsAltes:2005ph}.

The paper is organized as follows. In Sec. 2, we give the theoretical
background of our study and specify the observable we consider. In Sec. 3 we
present the numerical results of lattice Monte Carlo simulations.
Conclusions are given in Sec. 4.

%
 
\section{Theoretical setup}

The ultimate interest of our study is Euclidean pure SU($\Nc$) Yang-Mills theory, defined in continuum dimensional regularization by
\begin{equation} 
  \SE=\int \! \dd^dx \, \LE,  
  \quad\quad \LE=\frac{1}{2 g_3^2}\sum_{k,l} \Tr[F^2_{kl}] , 
\end{equation} 
where $d=3-2\epsilon$, $g_3^2$ is the gauge coupling, $k,l=1,\dots,d$,
$F_{kl}=i[D_k,D_l]$, $D_k=\partial_k-iA_k$, $A_k=A_k^aT^a$, 
and $T^a$ are Hermitean
generators of SU($\Nc$) normalized such that $\Tr[T^aT^b]=\delta^{ab}/2$.
The vacuum energy density in $\MSb$ (suppressing Faddev-Popov and
gauge fixing terms) is defined by 
\begin{equation}
  f_{\MSb}\equiv-\lim_{V\rightarrow\infty}
  \frac{1}{V}\ln\left[\int\!\mathcal{D}A_k\,
  \exp\left(-\SE\right)\right]_{\MSb},
\end{equation}
where $V$ denotes the $d$-dimensional volume. The use of the $\MSb$ dimensional
regularization scheme removes any $1/\epsilon$ poles from the
expression. In fact, using dimensional regularization the 
perturbative result vanishes, because there are no mass scales in the
propagators and therefore the UV and IR divergences cancel each
other. However, for dimensional reasons, the non-perturbative
form of the free energy is 
\begin{equation}
 f_{\MSb} = g_3^6 \left [A_G' \ln \frac{\bar{\mu}}{g_3^2} + B_G' \right],
\end{equation}
where $\bar{\mu}$ is the $\MSb$  renormalization scheme scale parameter.
The coefficient of the logarithm has been calculated by introducing a
mass scale $\mG^2$ for gluon and ghost propagators and sending
$\mG^2\rightarrow 0$ after the computation \cite{gsixg,sun}:
\begin{equation}
  f_{\MSb}=-g_3^6\frac{d_A\Nc^3}{(4\pi)^4}
 \left[\left(\frac{43}{12}-\frac{157}{768}\pi^2\right)
 \ln\frac{\bar{\mu}}{2\Nc g_3^2}
  + \BG(\Nc)+\mathcal{O}(\epsilon)\right],
 \label{MSb} 
\end{equation}
where $d_A = \Nc^2-1$. The non-perturbative constant part $\BG$, which
is a function of the number of colors, is what one would ultimately
like to determine. 

Using standard Wilson discretization, we can write the corresponding
action on the lattice as 
\begin{equation}
  S_a=\beta \sum_{\myvec{x}}\sum_{k<l}^3
  \left(1-\frac{1}{\Nc}\textrm{Re}\Tr[P_{kl}(\myvec{x})]\right),
\end{equation}
where $P_{kl}$ is the plaquette, $a$ is the lattice spacing and $\beta \equiv
2\Nc/(ag_3^2)$. Hence the continuum limit is taken by $\beta \rightarrow
\infty$. Analogously to $\MSb$, the free energy density is defined on the 
lattice as
\begin{equation}
  f_{a}\equiv-\lim_{V\rightarrow\infty}
  \frac{1}{V}\ln\left[\int\!\mathcal{D}U_k\,
  \exp\left(-S_a\right)\right].
\end{equation}
   Dimensionally, the vacuum energy density consists
of terms of the form $g_3^{2n}a^{n-3}$. Thus, approaching the continuum 
limit, we can relate $f_a$ and $f_{\MSb}$ as follows:
\begin{eqnarray}
  \Delta f & \equiv &  f_a-f_{\MSb} \label{deltaf}\\
  & = & C_1\frac{1}{a^3}\left(\ln\frac{1}{ag_3^2} +
  C_1'\right)+C_2\frac{g_3^2}{a^2}+C_3\frac{g_3^4}{a} + 
  C_4g_3^6\left(\ln\frac{1}{a\bar\mu}+C_4'\right)+\mathcal{O}(g_3^8a).
\end{eqnarray}

Taking derivatives of Eq.~(\ref{deltaf}) 
with respect to $g_3^2$ and using 3d
rotational and translational symmetries on the lattice, we obtain the
relation \cite{plaquette}  
\begin{equation}
  8\frac{d_A\Nc^6}{(4\pi)^4} \BG(\Nc) =\lim_{\beta
  \rightarrow\infty}\beta^4
  \left\{\ev{1-\frac{1}{\Nc}\Tr[P]}_a-
  \left[\frac{c_1}{\beta}+\frac{c_2}{\beta^2}
  +\frac{c_3}{\beta^3}
  +\frac{c_4}{\beta^4}(\ln\beta+c_4')\right]\right\}.
\label{master}
\end{equation}
The relations between $c_i$ and $C_i$ are 
\begin{align}
  c_1  &=  C_1/3
  &c_2 &=  -\frac{2\Nc}{3} C_2 
  &c_3  &=  -\frac{8\Nc^2}{3} C_3 \nonumber \\
  c_4  &=  -8\Nc^3C_4 
  &c_4' &=  C_4' -\frac{1}{3}-2\ln(2\Nc).
\end{align}
The first follows from a straightforward 1-loop computation:
\begin{equation}
  c_1=\frac{d_A}{3}.
\end{equation}
The 2-loop constant has been computed in three dimensions in \cite{hk} and can
be written as
\begin{eqnarray}
  c_2 & = & -\frac{2}{3}\frac{d_A\Nc^2}{(4\pi)^2}\left(\frac{4\pi^2}{3\Nc^2}+\frac{\Sigma^2}{4}
  - \pi\Sigma - \frac{\pi^2}{2} + 4\kappa_1+\frac{2}{3}\kappa_5\right)\\
      & = & d_A \Nc^2 \left(0.03327444(8)-\frac{1}{18}\frac{1}{\Nc^2}\right),
\end{eqnarray}
where the coefficients $\Sigma$, $\kappa_1$ and $\kappa_5$ can be found in \cite{farakos,lainerajantie}.
The 3-loop term has been computed in three dimensions recently in Ref.
\cite{pt}:
\begin{equation}
  c_3 = d_A\Nc^4\left(0.0147397(3) - 0.04289464(7)\frac{1}{\Nc^2} +
  0.04978944(1) \frac{1}{\Nc^4}\right).
\end{equation}
Because there is no $\bar{\mu}$ dependence in $f_a$, the value of $c_4$ is
determined by $f_{\MSb}$,
\begin{equation}
  c_4 = 0.000502301323 d_A\Nc^6 .
\end{equation}

The four-loop free energy itself is an IR divergent quantity at
in both $\MSb$ and lattice schemes. But the finite difference between
them, $c_4'$, can be defined by introducing the same IR cutoff, e.g. a
gluon mass, to both schemes. The cutoff dependence then cancels out
when the two schemes are compared. At present $c_4'$ is known only for
$\Nc=3$, for which it has been calculated using stochastic
perturbation theory \cite{renzo}.  

For later use we define the quantity 
\begin{equation}
  P_G(\beta,\Nc)\equiv \frac{32\pi^4 \beta^4}{d_A\Nc^6}
  \left\{\ev{1-\frac{1}{\Nc}\Tr[P]}_a-
  \left[\frac{c_1}{\beta}+\frac{c_2}{\beta^2}
  +\frac{c_3}{\beta^3}
  +\frac{c_4}{\beta^4}\ln\beta\right]\right\},
\end{equation}
which is a normalized plaquette expectation value minus all the
ultraviolet divergences. Hence, 
\begin{equation}
  \BG(\Nc) - \left(\frac{43}{12}-\frac{157}{768} \pi^2 \right)c_4'   =  P_G(\infty,\Nc).
\end{equation}
Our goal here is to determine $P_G(\infty,N_c)$. After the $\Nc$-dependence of $c_4'$ has been determined by, e.g., stochastic perturbation theory, one has reached the final goal, the determination of $B_G(N_c)$.

%

\section{Lattice computations}

The simulations were performed using Kennedy-Pendleton quasi heat bath (HB) \cite{kennedypendleton} 
and overrelaxation (OR) algorithms. For the overrelaxation we used an algorithm
which updates the whole matrix using singular value decomposition and
performs very well for large $\Nc$ \cite{forcrand}. Lattices of size
$N^3$, $N = 24,\ldots,400$ were used.  

For each HB update we performed one OR. The number of
updated subgroups in  HB for $\Nc=$ 3, 4, 5 and 8  were 3, 4, 8 and 24,
respectively. These subgroups were chosen randomly for each update.
 After each of these cycles we measured the
value of the plaquette. The integrated autocorrelation times were around
0.75. For SU(2) we used dedicated OR and HB algorithms, 
with a ratio of one OR step for each HB update. The autocorrelation
time was around 0.6. The data sets used for SU(3) are the same as in 
\cite{plaquette}. 

\begin{figure}  
  \centering
  \includegraphics*[width=\textwidth]{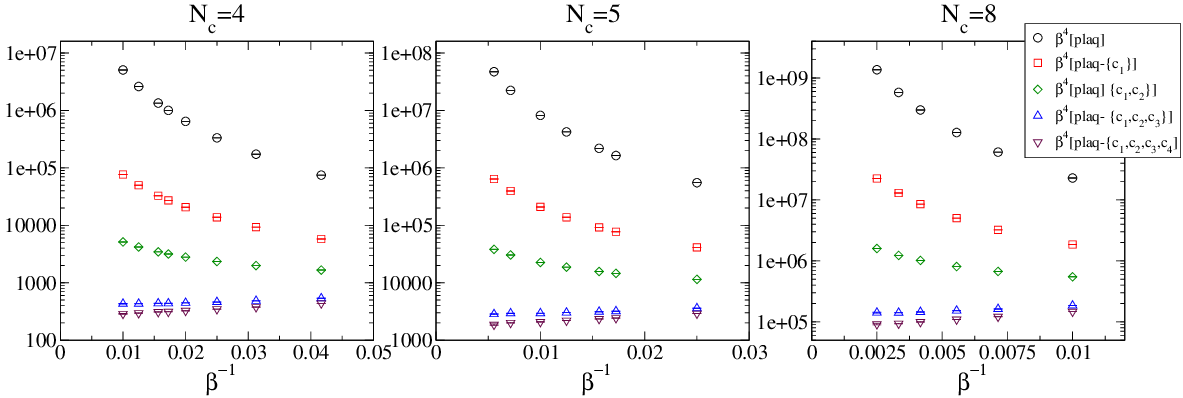}
  \caption{The significance loss due to the subtraction of ultraviolet divergences in the plaquette expectation value with different $\Nc$. Here ``plaq''$\equiv \ev{1-\frac{1}{\Nc} \Tr[P]}$ and the symbols $c_i$ in curly brackets represent which subtractions of Eq.~(\ref{master}) have been taken into account.}
  \label{sigloss} 
\end{figure}

The contribution of $B_G$ to the plaquette expectation value in
Eq. (\ref{master}) is about five orders of magnitude smaller than the
leading order contribution. Thus we experience massive significance
loss in the subtraction and the accuracy requirement makes the
numerical computation demanding (Fig.~\ref{sigloss}). 

The only physical scale in this problem is the correlation length of the lightest glueball, which according to \cite{teper} is $\sim 1/ \Nc g_3^2$. The requirement, that this scale be in the reach of the lattice gives us the condition
\begin{equation}
  a \ll \frac{1}{g_3^2\Nc} \ll  Na,
\end{equation}
which translates into
\begin{equation}
  2\Nc^2 \ll \beta \ll 2\Nc^2N.
\label{latticecondition}
\end{equation}

\begin{figure}
  \centering
\includegraphics*[width = 300pt]{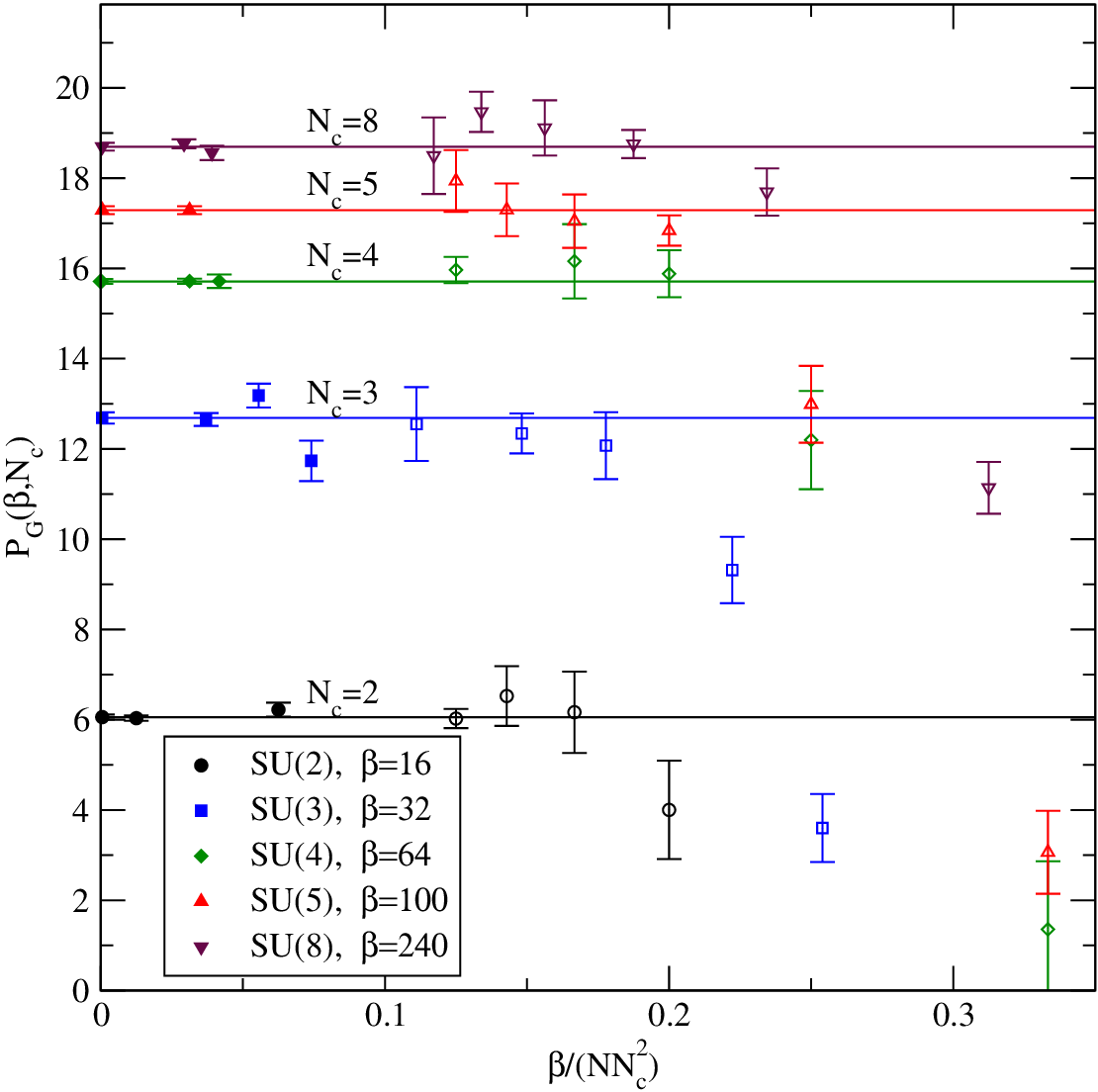}
  \caption{$P_G(\beta,\Nc)$ as a function
   of the physical lattice size $\beta/(N\Nc^2)$. Points denoted by
   open symbols are relatively low-statistics small volume
   simulations, included in order  to illustrate the exponentially
   suppressed finite volume effects. These  are omitted in the
   extrapolation. Finite-volume effects become visible when $
   \beta/(N\Nc^2) \sim 0.2$.  The points on the vertical axis indicate the
   infinite-volume estimate, obtained by fitting a constant to data in the
   range $\beta/(N\Nc^2)<0.1$.}  
  \label{finitevolume}
\end{figure}

Systematic errors due to the finite-volume effects turn out to be well under control.
Because the theory is confining, we expect finite-volume effects to be exponentially suppressed when the condition (\ref{latticecondition}) is fulfilled. As seen in Fig.~\ref{finitevolume}, the finite-volume effects are no longer visible within our resolution when $\beta \lesssim 0.2 \Nc^2 N$.

\begin{figure}[!ht]
  \centering
  \includegraphics*[width = 200pt]{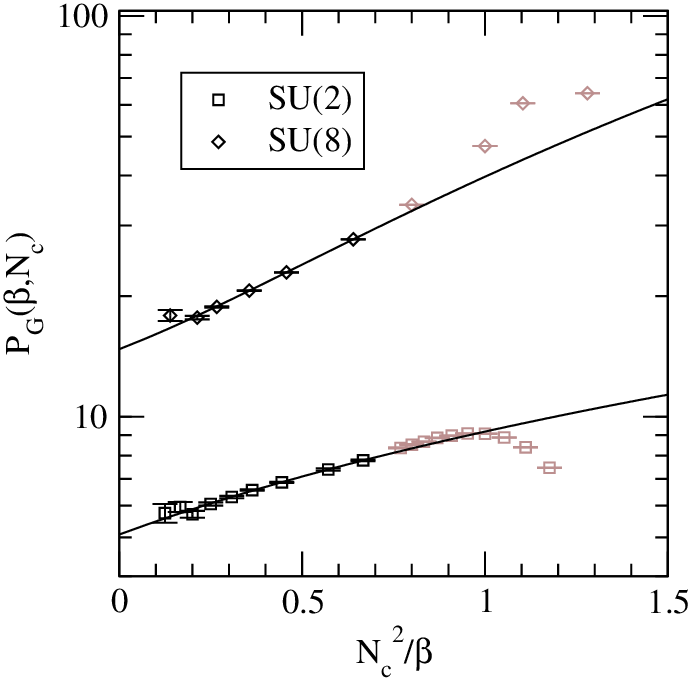}
  \caption{ The solid line indicates the continuum extrapolation obtained
    by fitting a second order polynomial to the infinite-volume extrapolated
    data. Points denoted by
    lighter color are omitted. The bulk phase transition point is
    around $\Nc^2/\beta \sim 0.9$. }
  \label{smallbeta} 
\end{figure}

In Fig.~\ref{smallbeta} the
effects arising from finite lattice spacing can be seen. We experience
a qualitative change in the  behavior of the plaquette expectation
value at $\beta \approx \Nc^2$. The plaquette expectation value as a
function of volume and lattice spacing $a$ is consistent with the
assumption of correlation lengths being $\sim \Nc^2/\beta$.

After numerous test runs we use in our simulations the requirement 
\begin{equation}
\Nc^2 < \beta \lesssim N(\Nc/3)^2,
\end{equation}
which is also the case in \cite{plaquette}.

\begin{figure}
\begin{center}
\includegraphics*[width = 250pt]{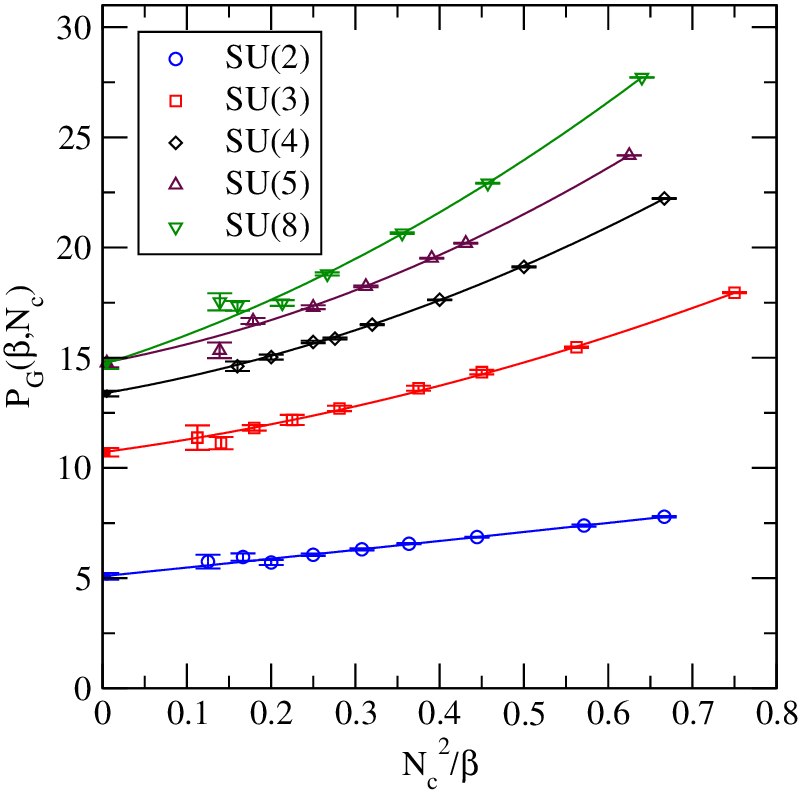}
\caption{Continuum extrapolations of infinite-volume extrapolated data for
  each $\Nc$. }
  \label{ext}
\end{center}
\end{figure}

The continuum extrapolation is obtained by fitting a polynomial
$P_G(\Nc)=d_1+d_2/\beta + d_3\beta^2$ to the infinite-volume
extrapolated data in Fig.~\ref{ext} for each $\Nc$ separately. This
functional form 
describes data quite well. The $\chi^2/$dof values for $\Nc=2,3,5$
are excellent but slightly discouraging for $\Nc=4,8$. The fitted values
are show in Table \ref{fit}. Using only
statistical errors of the fitting parameters would underestimate the
uncertainties of the continuum values, because the fit is dominated by
points far from the continuum limit. Inclusion of higher order terms
to the fitting function changes the continuum extrapolations by about
one sigma. Therefore we expect that the 1-sigma error of the continuum
extrapolated value is comparable to 2-sigma error of the fitting parameter
$d_1$.

\begin{figure}
  \centering
  \includegraphics*[width = 175pt]{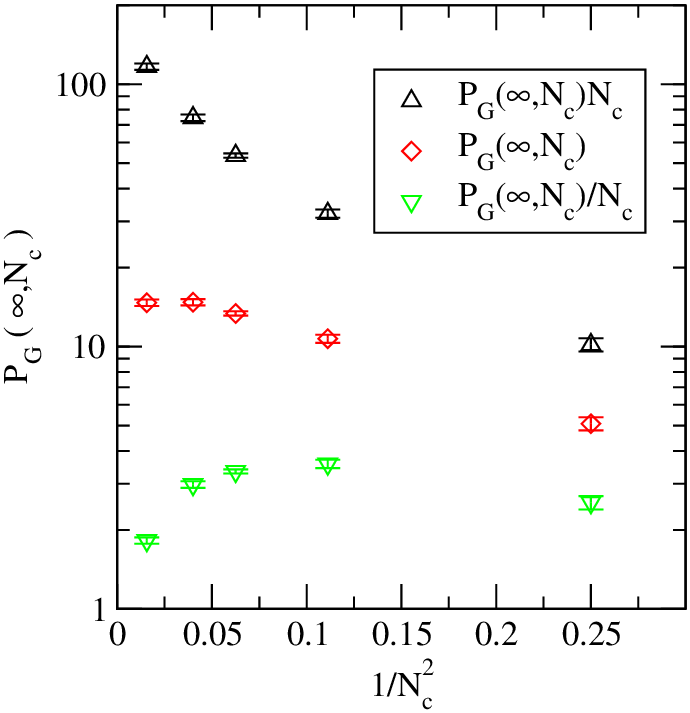}
  \caption{Comparing the leading order behavior of
  $P_G(\infty,\Nc)$ in $\Nc$. As predicted by planar theory, $P_G$
  approaches a constant in the large-$\Nc$ limit.  }
  \label{leadord}
\end{figure}

\begin{figure}
  \centering
  \includegraphics*[width = 200pt]{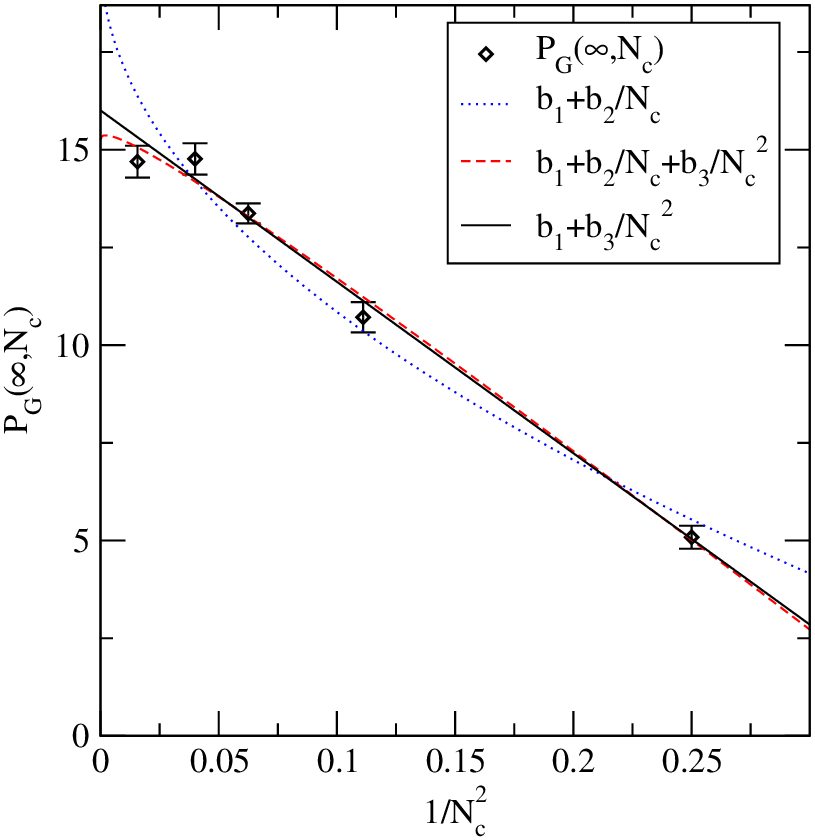}
   \caption{Comparing different fits for higher order terms in
  $\Nc$. The term $\Nc^{-1}$ is zero within our resolution implying
that $P_G$ is a function of $\Nc^{-2}$.}
  \label{higherord}
\end{figure}

At the leading order in $\Nc$, our measurements agree with the prediction of planar diagram
theory with $P_G(\Nc,\infty)$, approaching a constant (Fig.~\ref{leadord}). To study the
next order contributions we fit polynomials $b_1+b_2/\Nc$,
$b_1+b_2/\Nc+b_3/\Nc^2$ and $b_1+b_3/\Nc^2$ to the continuum
extrapolated data in Fig.~\ref{higherord}. We find that two last
forms fit the data quite well. The $b_2$ coefficient 
is zero (within our resolution) as could be expected from the form of the perturbative coefficients\footnote{Note, 
however, that terms $\sim 1/\Nc$ appear to be
possible in certain other pure gauge theory
observables~\cite{KorthalsAltes:2005ph}.},
which are also functions of $\Nc^2$. The data is
not accurate enough to determine higher order terms.

\begin{table}
  \centering
   \begin{tabular}{l|lcrcr|l||l}
     \hline
     \hline
     $\Nc$ & fit &&&&& $\chi^2/$dof & $P_G(\infty,\Nc)$ \\
     \hline
     \hline
     2 &  $5.09(15)   $&$+$&$ 16(3)      \beta^{-1} $&$   + $&$  3(11)\beta^{-2}$                  & $5.1/6$    & 5.1(3)  \\ 
     3 &  $10.7(2)    $&$+$&$ 46(7)      \beta^{-1} $&$   + $&$  4.85(6)\times 10^2\beta^{-2}$                 & $5.8/6$    & 10.7(4) \\ 
     4 &  $13.38(13)  $&$+$&$ 1.05(9) \times 10^2     \beta^{-1} $&$   + $&$  2.58(14)\times 10^3 \beta^{-2}$   & $12.3/5$   & 13.4(3) \\ 
     5 &  $14.8(2)    $&$+$&$ 1.8(2)\times 10^2    \beta^{-1} $&$   + $&$  7.9(5)\times 10^3   \beta^{-2} $  & $7.7/4$    & 14.8(4) \\ 
     8 &  $14.7(2)    $&$+$&$ 7.7(5) \times 10^2    \beta^{-1} $&$   + $&$  5.3(3)\times 10^4   \beta^{-2}$   & $17.7/4$    & 14.7(4) \\ 
     \hline
     \hline
   \end{tabular}
   \caption{The fitted values and $\chi^2/$dof of continuum extrapolations for
     each $\Nc$. The value in the brackets indicates the uncertainty of the
     last digit. The last column indicates the continuum limit with systematic errors included.}
   \label{fit}
\end{table}

As our final results we quote
\begin{equation}
 B_G(\Nc)+\left(\frac{43}{12}-\frac{157}{768}\pi^2\right)c_4'= P_G(\infty,\Nc)=15.9(2)-44(2)/\Nc^2
\end{equation}
Inserting $\Nc=3$ we get
\begin{equation}
  B_G(3)+\left(\frac{43}{12}-\frac{157}{768}\pi^2\right)c_4'= 11.0 \pm 0.3,
\end{equation}
which is consistent with the direct determination $10.7\pm 0.4$ \cite{plaquette}.

\begin{table}[b]
  \centering
  \begin{tabular}{l|l|l}
    \hline
    \hline
    function & values & $\chi^2/$dof \\
    \hline
    \hline
    $b_1+b_2\Nc^{-1}$             & $20.0(4)-28.9(12)\Nc^{-1}$                   &  $27.9/3$ \\
    $b_1+b_2\Nc^{-1}+b_3\Nc^{-2}$ & $15.25(11)+4.8(7)\Nc^{-1}-50.5(11)\Nc^{-2}$       &  $4.9/2$  \\
    $b_1+b_3\Nc^{-2}$             & $15.9(2)-43.5(17)\Nc^{-2}$                        &  $5.4/3$  \\
    \hline
    \hline
  \end{tabular}
  \caption{Different fitting functions for $P_G(\infty,\Nc)$. 
      The term $\Nc^{-1}$ provides a very bad description
      of the data (1st case) or has a coefficient consistent
      with zero within our resolution (2nd case); see also Fig.~\ref{higherord}. 
      The confidence values of fits are plausible for the last two functions.}
  \label{BGfit}
\end{table}

%

\section{Conclusions}   
The purpose of this paper has been to measure the $\Nc$-dependence of
the expectation value of the plaquette in three-dimensional pure gauge 
theory. We have also outlined how the continuum $\MSb$ scheme free energy
can be extracted from it. 
High precision lattice measurements of
plaquette were performed with $\Nc=2,3,4,5$ and $8$ and the large-$\Nc$ limit was taken by extrapolation. We found that the
non-perturbative input is $P_G = 15.9(2)-44(2)/\Nc^2$. 
The data does not seem to allow for terms $\sim 1/\Nc$,
and higher order terms, $\mathcal{O}(1/\Nc^3)$ or $\mathcal{O}(1/\Nc^4)$,
are small enough such that the physical case $\Nc = 3$
is very well described by this form. 

\section*{Acknowledgments}
We thank K. Rummukainen for his simulation code and useful
discussions. We also acknowledge useful discussions with K. Kajantie,
M. Laine and Y. Schr\"oder. This work was supported by the Magnus
Ehrnrooth Foundation, a Marie Curie Host Fellowship for Early Stage
Researchers Training, and Academy of Finland, contract numbers
104382 and 109720. Simulations were carried out at Finnish 
Center for Scientific Computing (CSC); the total amount of computing
power used was about $1.2\times10^{17}$ flops.

\appendix


\section{Tables}
In this Appendix we collect the numerical results for the plaquette
expectation value measurements, which have been used in the continuum 
extrapolations.
The column $N_\textrm{ind}$ gives the number of independent measurements within a data set. The data sets used for SU(3) are the same as in \cite{plaquette}. 

\begin{table}
  \begin{center}
  \begin{tabular}[t]{lr}
  \begin{tabular}[t]{|l|l|l|l|}
 \hline
   \multicolumn{4}{|c|}{SU(2)}\\
    \hline
    \hline
    $\beta$ & volume & $N_{\textrm{ind}}$ & $\ev{1-\frac{1}{\Nc}\Tr[P]}_a$ \\    
    \hline
    \hline 
    6   & $48^3$  & 40719  & 0.1752161(16) \\
    7   & $48^3$  & 42164  & 0.1488698(13) \\
    9   & $48^3$  & 43187  & 0.1145493(10) \\
    9   & $320^3$ & 5104   & 0.11454906(17) \\
    11  & $48^3$  & 42993  & 0.0931322(8) \\
    11  & $320^3$ & 8463   & 0.09313207(11) \\
    13  & $48^3$  & 44195  & 0.0784776(7) \\
    13  & $320^3$ & 8024   & 0.07847755(9) \\
    16  & $64^3$  & 157777 & 0.06350205(19) \\
    16  & $320^3$ & 14881  & 0.06350198(5) \\
    20  & $64^3$  & 271054 & 0.05062861(11) \\
    20  & $320^3$ & 12613  & 0.05062829(5) \\
    24  & $48^3$  & 904993 & 0.04209730(8) \\
    24  & $64^3$  & 317058 & 0.04209720(9) \\
    24  & $320^3$ & 14961  & 0.04209733(4) \\
    32  & $64^3$  & 868436 & 0.03148821(4) \\
    32  & $320^3$ & 15064  & 0.03148828(2) \\
    32  & $400^3$ & 6193   & 0.03148828(3) \\    
 \hline
 \end{tabular} 
&
 \begin{tabular}[t]{|l|l|l|l|}
   \hline
   \multicolumn{4}{|c|}{SU(3)}\\
   \hline
   \hline
   $\beta$ & volume & $N_{\textrm{ind}}$ & $\ev{1-\frac{1}{\Nc}\Tr[P]}_a$ \\ 
   \hline
   \hline  
12   & $24^3$ & 13459 & 0.2417125(8) \\
12   & $32^3$ & 10309 & 0.241717(6)  \\
12   & $48^3$ & 16236 & 0.241714(3)  \\
16   & $24^3$ & 15337 & 0.176526(6)  \\
16   & $32^3$ & 18668 & 0.176531(3)  \\
16   & $48^3$ & 19076 & 0.1765290(17)\\
16   & $64^3$ & 11833 & 0.1765302(14) \\
20   & $24^3$ & 11484 & 0.139295(5)  \\
20   & $32^3$ & 11634 & 0.139283(3)  \\
20   & $48^3$ & 19814 & 0.1392932(13)\\
24   & $24^3$ & 15992 & 0.115100(3)  \\
24   & $32^3$ & 20983 & 0.1151000(19)\\
24   & $48^3$ & 20723 & 0.1150986(11)\\
24   & $64^3$ & 12101 & 0.1151009(9) \\
32   & $48^3$ & 20451 & 0.0854789(8) \\
32   & $64^3$ & 24662 & 0.0854815(5) \\
32   & $96^3$ & 24875 & 0.0854806(3) \\
40   & $48^3$ & 20817 & 0.0680065(6) \\
40   & $64^3$ & 25442 & 0.0680058(4) \\
40   & $96^3$ & 25700 & 0.06800677(19)\\
50   & $64^3$ & 33448 & 0.0541741(3) \\
50   & $96^3$ & 69213 & 0.05417428(10)\\
50   & $128^3$& 29261 & 0.05417418(10)\\
50   & $320^3$& 8298  & 0.05417406(5)\\
64   & $96^3$ & 25211 & 0.04217128(12)\\
64   & $128^3$& 35565 & 0.04217113(6)\\
64   & $320^3$& 7921  & 0.04217123(4)\\
80   & $128^3$& 34310 & 0.03365240(6)\\
80   & $320^3$& 8356  & 0.03365247(3)\\
\hline
 \end{tabular}
 \end{tabular}
 \end{center}
\end{table}

\begin{table}
\begin{center}
 \begin{tabular}{lr}
   \begin{tabular}[t]{|l|l|l|l|}
     \hline
     \multicolumn{4}{|c|}{SU(4)}\\
     \hline
     \hline
     $\beta$ & volume & $N_{\textrm{ind}}$ & $\ev{1-\frac{1}{\Nc}\Tr[P]}_a$ \\
     \hline
     \hline  
     24  & $48^3$  & 79254 & 0.2257701(7) \\
     24  & $64^3$  & 15474 & 0.2257703(10) \\
     32  & $48^3$  & 58752 & 0.1651322(6) \\
     32  & $64^3$  & 16039 & 0.1651320(7) \\
     40  & $64^3$  & 16704 & 0.1303851(5) \\
     40  & $96^3$  & 33574 & 0.1303857(2)  \\ 
     40  & $128^3$ & 32872 & 0.13038581(13)\\
     50  & $64^3$  & 17257 & 0.1033093(4) \\
     50  & $96^3$  & 34295 & 0.10330876(16)\\ 
     50  & $128^3$ & 33813 & 0.10330879(10)\\
     58  & $96^3$  & 34810 & 0.08861313(14)\\ 
     58  & $128^3$ & 33443 & 0.08861289(9) \\   
     64  & $96^3$  & 17342 & 0.08007684(17)\\ 
     64  & $128^3$ & 50908 & 0.08007682(6) \\
     80  & $128^3$ & 50664 & 0.06372001(5) \\
     100 & $128^3$ & 51510 & 0.05076660(4) \\
     \hline
   \end{tabular}
&
 \begin{tabular}[t]{c}
   \begin{tabular}[t]{|l|l|l|l|}
     \hline
     \multicolumn{4}{|c|}{SU(5)}\\
     \hline
     \hline
     $\beta$ & volume & $N_{\textrm{ind}}$ & $\ev{1-\frac{1}{\Nc}\Tr[P]}_a$ \\      \hline
     \hline   40  & $128^3$ & 4667  & 0.2161236(5)\\
     58  & $128^3$ & 8515  & 0.1447591(2)\\
     64  & $128^3$ & 27137 & 0.13048507(12)\\
     80  & $128^3$ & 22616 & 0.10336875(10)\\
     100 & $128^3$ & 20238 & 0.08208926(8)\\
     140 & $128^3$ & 12886 & 0.05817279(8)\\
     140 & $160^3$ & 16049 & 0.05817267(5)\\
     180 & $128^3$ & 12184 & 0.04505589(6)\\
     180 & $160^3$ & 8597  & 0.04505586(5)\\
     \hline
  \end{tabular}
\\
 \begin{tabular}[t]{|l|l|l|l|}   
 \hline
   \multicolumn{4}{|c|}{SU(8)}\\
   \hline
   \hline
   $\beta$ & volume & $N_{\textrm{ind}}$ & $\ev{1-\frac{1}{\Nc}\Tr[P]}_a$ \\      \hline
   \hline
  100 & $96^3$ & 6493  & 0.2285506(5)\\
  140 & $96^3$ & 9789  & 0.1584135(3)\\
  180 & $96^3$ & 7127  & 0.1214678(2)\\
  180 & $128^3$ & 3522 & 0.1214678(2)\\
  240 & $96^3$ & 3755  & 0.0900746(3)\\
  240 & $128^3$ & 3857 & 0.09007497(16)\\
  300 & $96^3$ & 11266 & 0.07160377(11)\\
  300 & $128^3$ & 3831 & 0.07160353(13)\\
  400 & $96^3$ & 18120 & 0.05337892(6)\\
  400 & $128^3$ & 4251 & 0.05337902(8)\\
  460 & $128^3$ & 8656  & 0.04631022(5)\\
  \hline
 \end{tabular}
 \end{tabular}
 \end{tabular}
 \end{center}
\end{table}

\end{document}